# Group-Theoretic Partial Matrix Multiplication

Richard Strong Bowen, Bo Chen, Hendrik Orem,
and Martijn van Schaardenburg

Harvey Mudd College

**Abstract**

A generalization of recent group-theoretic matrix multiplication algorithms to an analogue of the theory of partial matrix multiplication is presented. We demonstrate that the added flexibility of this approach can in some cases improve upper bounds on the exponent of matrix multiplication yielded by group-theoretic full matrix multiplication. The group theory behind our partial matrix multiplication algorithms leads to the problem of maximizing a quantity representing the "fullness" of a given partial matrix pattern. This problem is shown to be $\mathbb{NP}$-hard, and two algorithms, one optimal and another non-optimal but polynomial-time, are given for solving it.

# Contents

# 1 Introduction

In 1969, Volker Strassen showed that the naïve algorithm for square matrix multiplication, which takes $O(n^3)$ time to multiply matrices of dimension $n$, is not optimal [8]; the algorithm he presented multiplied matrices in $O(n^{\log_2 7}) \approx O(n^{2.807})$. Together with the simple lower bound of $O(n^2)$ on the number of multiplications needed to multiply $n \times n$ matrices, Strassen's result originated the problem of determining the "best possible" exponent of matrix multiplication. To be precise, if $M(n)$ is the number of field operations in characteristic 0 required to multiply two $n \times n$ matrices, Strassen made the first step towards determining
$$\omega = \inf\{r \in \mathbb{R} | M(n) = O(n^r)\},$$
the exponent of matrix multiplication.

Gradual improvements were made to the upper bound on $\omega$. In 1990, Coppersmith and Winograd [4] showed that $\omega < 2.38$, a bound which remains the world record. A promising group-theoretic approach was presented by Cohn and Umans in 2003 [3]. They described an embedding of matrices into a group algebra that would allow for fast convolution via a Fourier transform, in much the same way that polynomials can be multiplied efficiently by embedding them in a group algebra, applying an FFT and then performing the convolution in the frequency domain. The challenge was to find an appropriate group together with three subsets which serve to index the matrix entries in the embedding. Using this method, Cohn et al. [2] tied the record of $\omega < 2.38$.

Proving a tight upper bound on $\omega$ is a long-standing open problem in theoretical computer science. It is widely believed that $\omega = 2$, but no progress has been made on the best known upper bound in nearly two decades.

In this paper, we generalize the results of Cohn et al., which only deal with full matrix multiplication, to a theory of group-theoretic partial matrix multiplication and use this approach to prove bounds on $\omega$. In particular, Theorem 2.12 states that
$$\omega \leq \frac{3\log\left(\sum_i d_i^\omega\right)}{\log f(A)},$$
where the $d_i$ are the character degrees of the chosen group and $f(A)$ represents, roughly, the amount of information computed in the product of two partial matrices of a particular "pattern."

The group-theory behind our partial matrix multiplication algorithm leads to an additional computational challenge, namely optimizing the quantity $f(A)$ given a set of possible patterns. We show this problem to be $\mathbb{NP}$-hard, and describe a non-optimal but polynomial-time algorithm, as well as an exponential-time algorithm for solving it. In a particular case, we show how to improve an upper bound on $\omega$ obtained in [2] by using the greater generality of group-theoretic partial matrix multiplication.

# 2 Full and Partial Group-Theoretic Matrix Multiplication

Our main theorems describe an algorithm for multiplying matrices using triples of subsets not satisfying the triple product property (see Definition 2.4). Some entries must be set to zero, and then partial matrix multiplications are performed. This section introduces the original group-theoretic algorithm by Cohn and Umans [3], as well as the notion of 'aliasing', the motivation for our focus on partial matrix multiplication.

## 2.1 Full Multiplication: The Cohn-Umans Algorithm

**Definition 2.1.** If $S, T, U$ are ordered subsets of a group $G$, then the Cohn-Umans algorithm [3] for matrix multiplication computes the product of matrices $M$ and $N$ of dimensions $|S| \times |T|$ and $|T| \times |U|$, respectively, as follows.

Index the rows of $M$ by $S^{-1}$, the columns of $M$ by $T$, the rows of $N$ by $T^{-1}$, and the columns of $N$ by $U$. Then let $f_M = \sum_{i,j} M_{i,j} s_i^{-1} t_j$ and $f_N = \sum_{j,k} N_{j,k} t_j^{-1} u_k$. Compute $f_P = f_M f_N$, and assign to $P_{i,k}$ the coefficient of $s_i^{-1} u_k$ in $f_P$.

**Theorem 2.2.** *The Cohn-Umans algorithm computes, in position $i, k$ of the product matrix, the sum of all terms $M_{i',j} N_{j',k'}$, where*

$$s_{i'}^{-1} t_j t_{j'}^{-1} u_{k'} = s_i^{-1} u_k.$$

*Proof.* Every term in $f_P$ is a product of a term in $f_M$ with a term in $f_N$. The $s_i^{-1} u_k$ term is exactly the sum of all terms $(zm)(z'n)$, where $z, z' \in \mathbb{C}^{n \times n}$, $m \in S^{-1}T$ and $n \in T^{-1}U$, and $mn = s_i^{-1} u_k$. But this is exactly the sum in the statement of the theorem. ∎

**Corollary 2.3.** *The Cohn-Umans algorithm is correct if and only if for all $s, s' \in S, t, t' \in T, u, u' \in U$, we have that $ss'^{-1} tt'^{-1} uu'^{-1} = e$ implies $s = s', t = t', u = u'$.*

*Proof.* This result follows from the previous theorem since

$$s_{i'}^{-1} t_j t_{j'}^{-1} u_{k'} = s_i^{-1} u_k$$

implies $i = i', j = j', u = u'$, meaning that entry $(i, k)$ of the product only contains terms formed by multiplying entry $(i, j)$ by $(j, k)$ in the left and right factor matrices, respectively. ∎

**Definition 2.4.** The property in 2.3 is called the *triple product property* [3].

**Example 2.5.** The following sets in $D_{12} = \langle x, y | x^6 = y^2 = 1, xy = yx^{-1} \rangle$ have the triple-product property:

$$S = \{1, y\}$$
$$T = \{1, yx^2, x^3, xy\}$$
$$U = \{1, yx\}$$

Thus, $S$, $T$, and $U$ can be used to index the product of a full $2 \times 4$ matrix by a ful $4 \times 2$ matrix, with no errors.

In this way, Cohn and Umans reduced the problem of proving bounds on $\omega$ to that of searching for groups with a good combination of character degrees and subsets satisfying the triple product property. It is, however, unnecessary to require that the group element index sets produce a fully correct product. Even when terms in the group algebra multiplication incorrectly appear in an entry of the product matrix due to a violation of the triple product property by our chosen subsets $S, T$, and $U$ (we call this phenomenon *aliasing* to emphasize the analogy to the usual Fourier transform in signal processing), these index sets will still compute the correct product in the case where one of the input entries contributing to each aliasing term contains a zero.

In the next section, we show how to apply the classical theory of partial matrix multiplication to the group-theoretic framework developed by Cohn et al. We will present bounds on $\omega$ realizable through subsets which may or may not satisfy the triple product property; in a special case, we can show that our algorithm yields strictly stronger results than the original Cohn-Umans full matrix multiplication algorithm. For a specific family of constructions satisfying the triple product property, the associated bound on $\omega$ can be improved by adding a single element to each of the sets, described in Section 4. This means that the additional information computed by increasing the matrix dimensions outwiegehs the information lost due to the partial nature of the larger multiplication.

## 2.2 Partial Multiplication: Aliasing

**Definition 2.6.** If $S, T, U$ are subsets of a group $G$, the set of all triples $((i, j), (j', k), (i', k'))$ where

$$s_i^{-1} t_j t_{j'}^{-1} u_k = s_{i'}^{-1} u_{k'}$$

and $i \neq i', j \neq j'$, or $k \neq k'$ is called the *set of aliasing triples*, $A$.

Aliasing sets can be visualized as sets of lines representing the triples as shown in Figure 1. Each line is broken up into two pieces: the first runs from the left factor matrix to the right factor matrix and represents which pair of input entries combine to produce an incorrect term in the product; the second runs from the right factor matrix to the product, indicating where the incorrect term appears.

**Definition 2.7.** The *left aliasing set* of a set of aliasing triples $A$ is

$$\{x : \text{there exist } y, z \text{ such that } (x, y, z) \in A\}.$$

The right aliasing set and the product aliasing set are defined analagously. The left aliasing set is the set of indices in the left factor matrix in Figure 1 that are the endpoints of one of the lines.

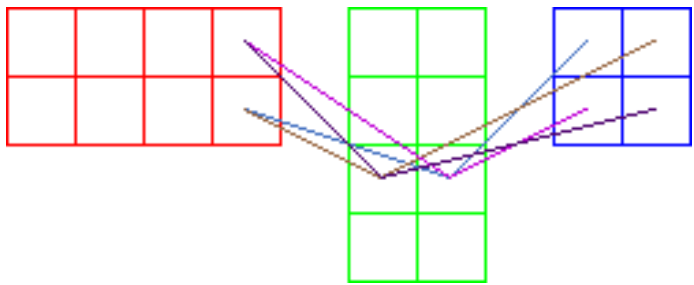

Figure 1: A visualization of the aliasing set in Example 2.10, where the input matrices are the left and middle rectangles, and the output is on the right. A triple $((i,j),(j',k),(i',k'))$ corresponds to a pair of lines from $(i,j)$ in the left factor matrix to $(j',k)$ in the right, and from $(j',k)$ in the right factor matrix to $(i',k')$ in the product; the set of all $(i,j)$ which is the start of a line in the diagram is the left aliasing set.

It is impossible to have only one of $i \neq i', j \neq j', k \neq k'$ (if, for example, only $i \neq i'$ held, then we would have $s_i^{-1} e u_k = s_{i'}^{-1} u_k$). Thus, an incorrect term in the Cohn-Umans algorithm will only occur having at least two of

1. being in the wrong row given its first multiplicand,
2. being in the wrong column given its second multiplicand, or
3. having its multiplicands coming from different positions in their respective row and column.

**Definition 2.8.** Let $A$ be a set of aliasing triples for $S, T, U \subseteq G$. We say that $I$ and $J$ *cover* $A$ if $I$ and $J$ are subsets of the indices of entries of a $|S| \times |T|$ and $|T| \times |U|$ matrix, respectively, such that for all $a$ in $A$, either the first entry of $a$ is in $I$ or the second is in $J$. If $M$ and $N$ are $|S| \times |T|$ and $|T| \times |U|$ entries such that for every index $i$ in $I$, $M_i$ is 0, and similarly for $N$ and $J$, we say that $M$ and $N$ *realize* $I$ and $J$.

**Theorem 2.9.** *Let $G$ be a group and let $S, T, U$ be indexing sets with aliasing set $A$. Let $M, N$ be matrices of size $|S| \times |T|, |T| \times |U|$, respectively, and let $I, J$ be subsets of the indices that cover $A$. If $M, N$ realize $I, J$, then the Cohn-Umans algorithm correctly computes the partial matrix product $MN$.*

*Proof.* By Theorem 2.2, the extra terms arise from entries in the input matrices with indices in the aliasing set $A$. Thus setting the entries corresponding to entries of $I$ and $J$ to zero sets the coefficient on each incorrect term to zero, yielding the correct product of the partial matrices of $M, N$. ∎

**Example 2.10.** Consider our earlier example in $D_{12}$, with a change to the last element of $T$:

$$S = \{1, y\}$$
$$T = \{1, yx^2, x^3, x^4\}$$
$$U = \{1, yx\}.$$

This triple has aliasing set

$$\begin{aligned}
A = \{&((2,4), (3,2), (1,1)), \\
&((2,4), (3,1), (1,2)), \\
&((1,4), (3,2), (2,1)), \\
&((1,4), (3,1), (2,2))\},
\end{aligned}$$

as depicted in Figure 1. The first element of $A$ describes the indices in the product

$$s_2^{-1} t_4 t_3^{-1} u_2 = s_1^{-1} u_1$$

that erroneously form an extra term in the top left corner of the product matrix. Thus, using these sets, the Cohn-Umans algorithm correctly computes these types of partial matrix multiplication:

$$\begin{bmatrix} a_{1,1} & a_{1,2} & a_{1,3} & 0 \\ a_{2,1} & a_{2,2} & a_{2,3} & 0 \end{bmatrix} \times \begin{bmatrix} b_{1,1} & b_{1,2} \\ b_{2,1} & b_{2,2} \\ b_{3,1} & b_{3,2} \\ b_{4,1} & b_{4,2} \end{bmatrix}$$

$$\begin{bmatrix} a_{1,1} & a_{1,2} & a_{1,3} & a_{1,4} \\ a_{2,1} & a_{2,2} & a_{2,3} & a_{2,4} \end{bmatrix} \times \begin{bmatrix} b_{1,1} & b_{1,2} \\ b_{2,1} & b_{2,2} \\ 0 & 0 \\ b_{4,1} & b_{4,2} \end{bmatrix}.$$

The aliasing triples are visually depicted in Figure 1.

We will now introduce a function that will be an integral part of our partial matrix multiplication algorithm. It computes the *number of ones in a tensor of partial matrix multiplication*, which intuitively means the amount of information computed by this partial multiplication. Its importance will become clear in the next theorem.

**Definition 2.11.** Let $A$ be a set of aliasing triples and let $I$ and $J$ cover $A$. The function $f(I, J)$ is equal to

$$\sum_i k_i n_i,$$

where $k_i$ is the number of entires in the $i^{th}$ column of the left factor matrix which do not appear in $I$ and $n_i$ is the number of entries in the $i^{th}$ row of the right factor matrix which do not appear in $J$. Finally, $f(A)$ is

$$f(A) = \max\{f(I, J) | I \text{ and } J \text{ cover } A\}.$$

The function $f$ is a measure of how much computation is being done by a partial matrix multiplication. Notice that if there is no zeroing in a multiplication of $m$ by $n$ by $n$ by $p$, then by $I$ and $J$ both empty, $f \geq mnp$ (and it's easy to see that $f = mnp$). The following theorem is used to derive many of our results; it provides a bound on $\omega$ given subsets which need not satisfy the triple product property. For this proof, it is sufficient to consider only matrices of complex numbers. Note that in the special case where the aliasing set is empty (that is, $S, T, U$ have the triple product property), $f(A) = |S||T||U|$ and our bound recovers Theorem 1.8 in [2]. This mimics the proof of Theorem 4.1 in [3], and uses some if its terminology.

**Theorem 2.12.** *Let $S, T, U \subseteq G$ with aliasing triples $A$, and suppose $G$ has character degrees $\{d_i\}$. Then*

$$\omega \leq \frac{3 \log(\sum_i d_i^\omega)}{\log f(A)}$$

*Proof.* Let $t$ be the tensor of partial matrix multiplication corresponding to $I, J$, the patterns which maximize $f$. It is clear that

$$t \leq \mathbb{C}G \cong \bigoplus_i \langle d_i, d_i, d_i \rangle$$

(similar to Theorem 2.3 in [3]). Then the $l^{th}$ tensor power of $t$ satisfies

$$t^l \leq \bigoplus_{i_1, \ldots, i_l} \langle d_{i_1} \ldots d_{i_l}, d_{i_1} \ldots d_{i_l}, d_{i_1} \ldots d_{i_l} \rangle.$$

By the definition of $\omega$, each $\langle d_{i_1} \ldots d_{i_l}, d_{i_1} \ldots d_{i_l}, d_{i_1} \ldots d_{i_l} \rangle$ has rank at most $C(d_{i_1} \ldots d_{i_l})^{\omega+\varepsilon}$ for some $C$ and for all $\varepsilon$. So, taking the rank of both sides gives

$$R(t)^l \leq C \left( \sum d_i^{\omega+\varepsilon} \right)^l,$$

from Proposition 15.1 in [1]. Since this is true of all $\varepsilon > 0$, it holds for $\varepsilon = 0$ by continuity:

$$R(t)^l \leq C \left( \sum d_i^\omega \right)^l.$$

Taking $l^{th}$ roots as $l \to \infty$ gives

$$R(t) \leq \sum_i d_i^\omega.$$

By Theorem 4.1 in [7]

$$\omega \leq \frac{3 \log(\sum_i d_i^\omega)}{\log f(A)}.$$

∎

# 3 Algorithms for Aliasing Structure

In the study of aliasing, the following problem comes up: there is a pattern $A$, and one wishes to find the value $f(A)$ by trying various $I, J$. This problem is $\mathbb{NP}$-hard; this section describes the worst-case exponential-time algorithm we use to solve it exactly, as well as a polynomial time algorithm used to find a reasonable solution.

## 3.1 A Polynomial-Time Non-Optimal Algorithm for Finding Aliasing Covers

In this section we will give a polynomial-time algorithm for finding covering sets $I, J$. This is not an approximation algorithm in the complexity-theoretic sense; it is merely a "pretty good" algorithm which we found useful in research. Instead of finding the cover which minimizes $f$, we find the cover which zeros the fewest entries. Viewing the entries in the factor matrices as vertices in a bipartite graph, and the pairs in the aliasing set as edges, it is clear that we desire a minimal vertex cover. By König's theorem, this is equivalent to finding a maximum matching (for an excellent explanation of the associated algorithm, see [5]), which can be solved efficiently in bipartite graphs with [6].

## 3.2 Computing the Optimal Cover for Aliasing

When computing $f$ by exhaustive search, one must choose, for each aliasing triple, whether to satisfy it by zeroing the left or by zeroing the right. After each choice, however, one can compute the current value of $f$ as if the only triples in $A$ were those already assigned a zero. Then making further choices will only lower this value of $f$, so if the computed value is below the already known best value, the entire search tree can be pruned. In pseudocode,

```
procedure maximum_f(A)
  S = new Stack
  F = new Frame(A) #meaning that F stores A, the set of aliasing
    triples; and I and J, the trial patterns, currently empty
  bestf = -1
  bestfFrame = F
  while S is not empty
    frame = S.pop()
    if every triple in A is covered by frame.I and frame.J and
        f(frame.I,frame.J) > bestf then
        bestf = f(frame.I,frame.J)
        bestfFrame = F
        continue
    if f(frame.I,frame.J) <= bestf then continue #don't need this subtree
    a = first triple in A not covered by frame.I, frame.J
    frame1 = copy(frame)
```

```
    frame2 = copy(frame)
    frame1.I.append(left entry of a)
    frame2.J.append(right entry of a)
    S.push(frame1,frame2)
```

# 4 Improving a Group-Theoretic Construction through Aliasing

In this section we present an improvement over a construction presented in §2 of [2].

## 4.1 The Original Construction

Let
$$H = C_n \times C_n \times C_n,$$
$$G = H \wr S_2,$$

and let $H_i < H$ be the subgroup isomorphic to $C_n$ in the $i$th coordinate. By $z$ we mean the generator of $S_2$, and by $e_H$ we mean the identity element of $H$.

We write elements of $G$ as
$$(a,b)z^j$$
where $a, b \in H$ and $j = 0$ or $1$.

Define, for $i \in 1, 2, 3$, the subsets of $G$
$$S_i = \{(a,b)z^j | a \in H_i \setminus e_H, b \in H_{i+1}, j = 0 \text{ or } 1\}$$

where subscripts are taken mod 3. Finally, we let
$$S = S_1, T = S_2, U = S_3.$$

By [2], Lemma 2.1, $S, T, U$ have the triple product property. Note that
$$|S| = |T| = |U| = 2n(n-1),$$
and so
$$f = 8n^3(n-1)^3.$$

This construction gives $\omega \leq 2.9088$ for $n = 17$.

## 4.2 Relaxing the Triple Product Property

Let $S_i$ be as defined in the previous section, and let
$$S_i' = S_i \cup \{(e_H, e_H)\}.$$

Let $S' = S_1', T' = S_2', U' = S_3'$, and let $A$ be the associated aliasing set, shown graphically in Figure 2.

We find that $A$ can be partitioned into three easily analyzed categories:

(a) "Bottom aliasing" occurs in the rows of the product which are not indexed by the identity. All aliasing of this type can be covered by zeroing some $(n-1)^2$ entries in the $(e_H, e_H)$ column of $R$.

(b) "Top-Easy aliasing" occurs in the $(e_H, e_H)$ row of the product. These are entirely covered by zeroing $(n-1)^2$ entries of the $(e_H, e_H)$ column of $L$.

(c) "Top-Hard aliasing" also occurs in the $(e_H, e_H)$ row of the product. The distinction is in the manner in which they arise. Alasing in this category can be covered by two things: the same entries which cover Top-Easy aliasing, combined with an additional $2n(n-1)$ entries in the $(e_H, e_H)$ column of $R$.

This decomposition is depicted in Figure 3.

There exists a pair $I, J$ with $(n-1)^2$ elements in the first column in $L$, and the entire first column in $R$, that cover $A$. Thus

$$f \geq (2n(n-1))^3 + (2n(n-1))^2 + (2n(n-1))\big[2n(n-1) - (n-1)^2 + 1\big],$$

which is strictly greater than $f$ for $S, T, U$. For $n = 17$, we acheive $\omega \leq 2.9084$.

The insight here is that we only zeroed entries that we added. That is, this partial matrix multiplication contains the entire matrix multiplication indexed by $S, T, U$, and then some more. Thus, by relaxing the restriction on $S, T, U$, we *strictly increased* the amount of computation done, without increasing the work necessary (since $G$ is constant).

## 5  The Complexity of Computing a Best Cover

Often we are confronted with this problem: given some triple of subsets, find the best way to put aliasing in the factor matrices and have the best bound on $\omega$, i.e., the best $f(I, J)$. We show this problem is computationally hard.

Consider the problem PARTIAL-TENSOR-ONES: given the dimensions of two matrices $m, n, p$, a set of pairs $A = \{((a_i, b_i), (c_i, d_i))\}$, and an integer $k$, are there $I$ and $J$ realizing $A$ such that $f(I, J) = k$? (This is the problem of maximizing the dot product when all the aliasing is to be taken care of by the left and right matrices). We show that PARTIAL-TENSOR-ONES is $\mathbb{NP}$-complete via a reduction from INDEPENDENT-SET, a well-known $\mathbb{NP}$-complete problem.

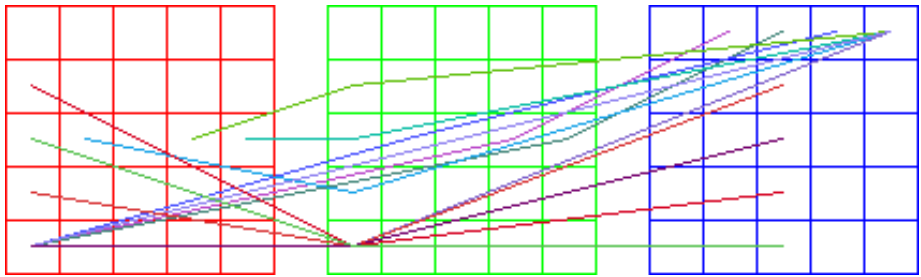

Figure 2: A visualization of the aliasing in the construction introduced in Section 4.2. In this case, $n = 2$.

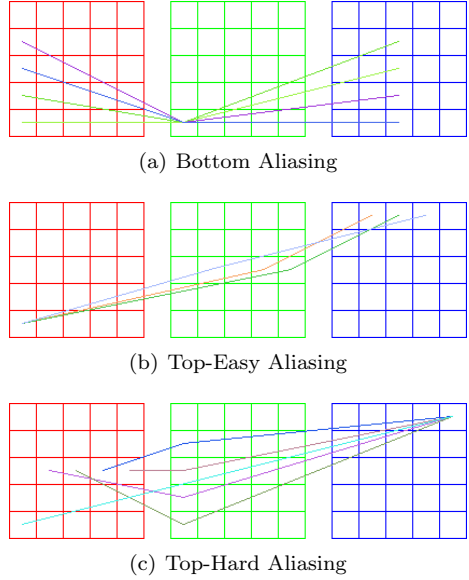

(a) Bottom Aliasing

(b) Top-Easy Aliasing

(c) Top-Hard Aliasing

Figure 3: A visualization of the three types of aliasing in the construction given in Section 4.2.

**Theorem 5.1.** *PARTIAL-TENSOR-ONES is $\mathbb{NP}$-complete*

*Proof.* An instance of INDEPENDENT-SET consists of (some encoding of) a graph and an integer $k$. Let $G = (V, E)$ be this graph. We will generate an instance of PARTIAL-TENSOR-ONES. Let $m = p = |V|$ and $n = 1$. For each edge $(v_i, v_j)$, add constraints of the form $((1, i), (j, 1))$ and $((j, 1), (1, i))$.

Suppose there is an independent set of size $k$. Then there is an $I, J$ such that $f(I, J) = k$. For each $v_i$ in the independent set, allow $(i, 1)$ and $(1, i)$ to be free and all other entries in the two vectors to be zeroed. It's clear that $f(I, J) = k$, and every constraint is fulfilled because the constraints correspond exactly to the edges, so no two free variables appear in the same constraint.

From an aliasing pattern with $f(I, J) = k$, we can construct an independent set of the same size. If any $(1, i)$ is free in $I$ while $(i, 0)$ is zeroed in $J$, modify $I$ to set $(1, i)$ to zero. Then the value of $f$ is unchanged, but all pairs are either both free or both 0. This is the sort of aliasing pattern one gets from the previous reduction, and we can easily run the argument of the previous paragraph backwards to find an independent set in $G$ of size $k$.

Since there is an independent set of size $k$ if and only if there are some $I, J$ such that $f(I, J) = k$, and the reduction is clearly polynomial time, PARTIAL-TENSOR-ONES is $\mathbb{NP}$-hard.

To show that PARTIAL-TENSOR-ONES is in $\mathbb{NP}$, we must show a polynomial-sized certificate which can be checked in polynomial time. Given an instance of PARTIAL-TENSOR-ONES, a certificate can be a list of three symbols, $L$, $R$, or $B$, one for each constraint, indicating whether that constraint is satisfied by zeroing on the left, on the right, or in both. This is clearly polynomial in size of the input. To check the certificate, one only needs to check two conditions: first, that it is consistant, that is, that no pair of constraints on the same entry of the matrix constrain it to be both free and zero, which can be done with the square of the number of constraints such checks, and second that $f(I, J) \geq k$, which can be done by making a list of rows and columns with zeored entries, and for each of these the number of nonfree entries in that row or column. Then $f$ can be computed from this easily. This takes time proportional to the number of constraints as well. So, the certificate can be verified in polynomial time. Therefore, PARTIAL-TENSOR-ONES is $\mathbb{NP}$-complete. ∎

*Remark*: We have not shown, in the reduction, a *group* (and appropriate subsets) which provides the appropriate aliasing. So, any polynomial time algorithm to find the best aliasing pattern from a given group and triple of subsets must either use more group theory, or show that $\mathbb{P} = \mathbb{NP}$.

# 6 Conclusion

We have shown that an analogue the algorithm described in [3] can be applied to indexing sets that do not satisfy the triple product property, and provide some techniques for addressing the resulting optimization problems. In particular, we take sets satisfying the property and modify them in a small way to achieve a lower bound on $\omega$. As the group-theoretic approach is known to tie the best-known upper bound, this suggests a possible path to improving upon the current record.

# 7 Acknowledgements

We would like to thank our research advisors, Professors Michael Orrison and Nicholas Pippenger of Harvey Mudd College Mathematics, for the opportunity to work on this project, as well as their assistance and collaborative efforts throughout the summer. Professor Chris Umans of Caltech Computer Science, was also very generous with his time and supportive of our research efforts.

We are grateful for the help of Claire Connelly in setting up our computing environment and allowing us to use `hex`, a parallel-computing system operated by Harvey Mudd College's Department of Mathematics and the Quantitative Life Sciences Center. `hex` was made available through a grant from the W.M. Keck Foundation.

Our research was funded in part by National Science Foundation grant CCF 0430656.